\documentclass[11pt,preprint]{aastex}

\shorttitle{A Flattened Protostellar Envelope}
\shortauthors{Looney et al.}

\begin{document}

\title{A Flattened Protostellar Envelope in Absorption around L1157}
\author{Leslie W. Looney\altaffilmark{1}, John J. Tobin\altaffilmark{2}, and Woojin Kwon\altaffilmark{1}}

\altaffiltext{1}{Department of Astronomy,
University of Illinois, 1002 W. Green St., Urbana, IL 61801, lwl@uiuc.edu, wkwon@uiuc.edu}
\altaffiltext{2}{Department of Astronomy, University of Michigan, 500 Church St., Ann Arbor, MI 48108, jjtobin@umich.edu}      

\begin{abstract}

Deep \textit{Spitzer} IRAC images of L1157 reveal many of the details of
the outflow and the circumstellar environment of this Class 0 protostar. In IRAC band 4, 8 $\mu$m, 
there is a flattened structure seen in absorption against the background emission. 
The structure is perpendicular to the outflow and is extended to a diameter of $\sim$2$\arcmin$.
This structure is the first clear detection of
a flattened circumstellar envelope or pseudo-disk around a Class 0 protostar.
Such a flattened morphology is an expected outcome for many collapse theories that
include magnetic fields or rotation.
We construct an extinction model
for a power-law density profile, but we do not constrain the density power-law index.
\end{abstract}


\section{Introduction}
\label{dist}

The L1157 dark cloud in Cepheus (IRAS 20386+6751) conceals a young
protostar, a so-called Class 0 source \citep{andre1993}, which is deeply
embedded within a large circumstellar envelope \citep{gueth2003,beltran2004}.  L1157 has 
a large powerful molecular outflow that is the prototype of chemically active outflows
\citep{bachiller2001}.
Despite the attention that L1157 has received at radio wavelengths,
few observations have been made in the near to mid-infrared outside
of observations of the outflow in H$_2$ and K-band \citep[e.g.][]{davis1995,cabrit1998}.
Only recently have sensitive instruments been available to observe
these objects shortward of 10 $\mu$m \citep[e.g.,][]{tobin2007}.
The outflow carves cavities in the
circumstellar envelope, which allow photons from the embedded central
source to escape and scatter off dust in the cavity at NIR wavelengths.
The morphology of the scattered light can be used to probe many of the
fundamental properties of the source such as opening angle, envelope mass,
etc. \citep[e.g.,][]{whitney2003a,whitney2003b,tobin2007,robitaille2007,seale2008}.

In this letter, we present new, deep \textit{Spitzer Space Telescope}
observations of L1157.
The IRAC continuum
emission is dominated by molecular line emission in the outflow.
Near the source there is a small amount of
emission that may be attributed to scattered light and perhaps molecular line emission
that is highly excited by the outflow jet. Other than the enormous
outflow ($\sim$0.5 pc) to the north and south, the most prominent feature
observed is a large, flattened absorption feature
at 8.0 $\mu$m and less defined at 5.8 $\mu$m. This absorption feature
is a flattened circumstellar envelope observed in silhouette against
the Galactic infrared background.

The distance to L1157 is important to any physical interpretation, but the
distance is highly uncertain. The molecular clouds in Cepheus have three
characteristic distances, 200, 300, 450 pc \citep{kun1998}; L1157 has
a similar galactic latitude as the 200 pc and 300 pc absorbing clouds.
Due to this, we disagree with the current accepted distance of 440
pc. This value was based upon a study of NGC 7023 in \citet{viotti1969};
L1157 is not in clear association with this cluster.  In this letter,
we use a distance of 250 pc.

\section{Observations}

The biggest breakthrough in observing Class 0 sources in scattered
light has come with the sensitivity of the Infrared Array Camera
(IRAC) \citep{fazio2004} on the \textit{Spitzer Space Telescope}
\citep{werner2004}.  This enables observers to see through the dust
enshrouding a Class 0 source and reveal structures close to the embedded
source via scattered light and emission lines.  IRAC has channels numbered
1 through 4, corresponding to central wavelengths of 3.6 $\mu$m, 4.5
$\mu$m, 5.8 $\mu$m, and 8.0 $\mu$m, respectively.

\begin{figure}
\includegraphics{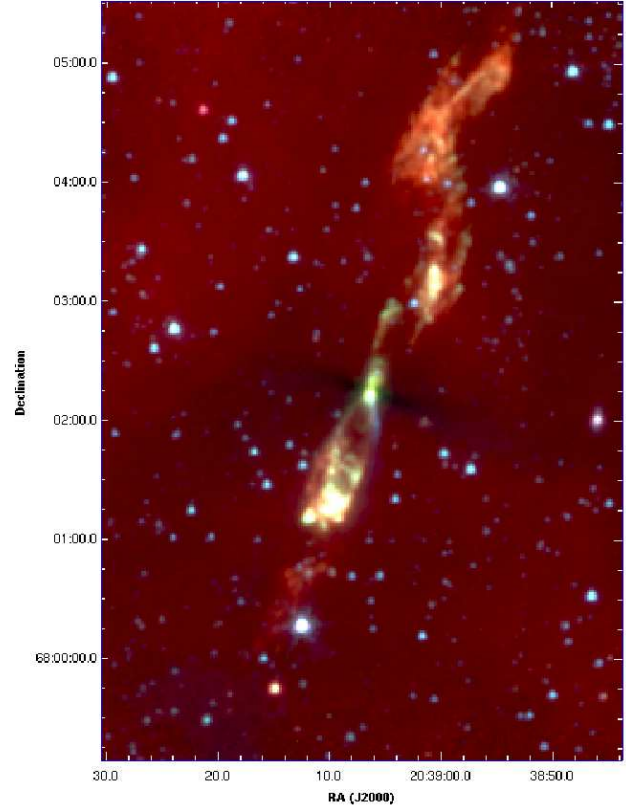}
\vspace{8cm}
\caption{Color IRAC image of the L1157 region with Ch1-blue, Ch2-green, and Ch4-red.  
The color stretch is slightly exaggerated to emphasize Ch4 (8 $\mu$m band)
where the extinction is the largest.}
\label{color}
\end{figure}

\begin{figure}
\includegraphics{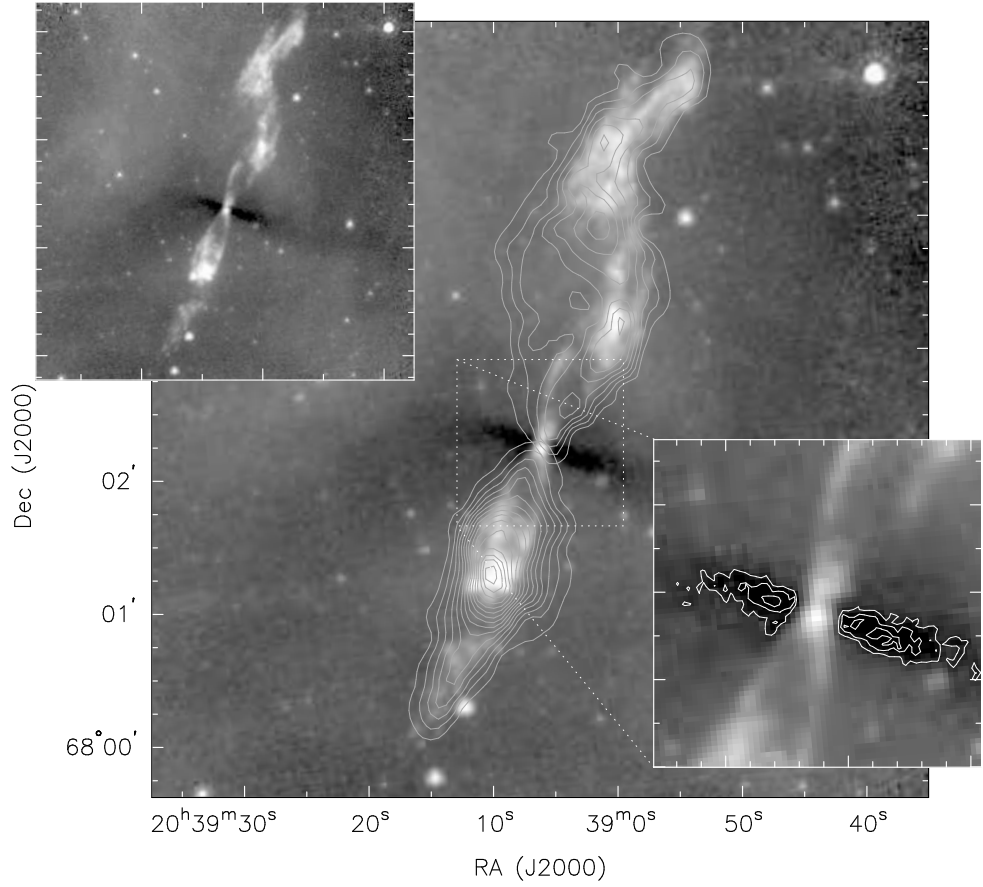}
\vspace{8cm}
\caption{IRAC band 4 greyscale image of L1157 overlaid with the CO 2-1 emission
from \cite{bachiller2001}.  The upper-left inset is
the same greyscale without the contours.
The lower-right inset is a closeup ($\pm$75$\arcsec$) of the absorption structure
with contours of 0.042 MJy/sr $\times$ 1, 2, 3, and 4.  
}
\label{disk}
\end{figure}

The L1157 cloud was observed with IRAC on 2006 August 13.  Observations
were carried out using the High Dynamic Range mode with frame times
of 30 seconds using a cycled dither pattern of 30 positions and a
small scale factor achieving a total integration time of 900 seconds.
This observing scheme increases the overall sensitivity to scattered
light in the outflow cavity of Class 0 sources \citep[e.g.,][]{tobin2007}.
Post-BCD pipeline products (version S14.4.0) were solely used in this letter.  
A color-composite image, with
exaggerated band 4 stretch of the field is shown in Figure \ref{color}.
In addition, a greyscale image using only IRAC channel 4, 8 $\mu$m band,
is shown in Figure \ref{disk}.

\section{Results}

\subsection{Flattened Envelope in Absorption}

Even a cursory examination of Figures \ref{color} and \ref{disk}
reveals a clear absorption feature perpendicular to the outflow.
Much like the proplyds in Orion \citep[e.g.,][]{odell1993}, the
structure is seen in absorption against a bright background.  In our
case, the
bright and pervasive background emission is from the 8.0 $\mu$m band.
There is also a lower level of intrinsic background emission in the 5.8
$\mu$m channel, which shows a less prominent absorption feature. However,
the level of absorption is too low for a rigorous analysis.
If the absorption is due to the opacity of the circumstellar material,
we would also expect to see absorption features in the 3.6 and 4.5 $\mu$m,
but in those channels, the background radiation field is nearly zero.
The most likely explanation is that the background emission is from Polycyclic Aromatic
Hydrocarbon (PAH) features, which are strongest in the 5.8 and 8 $\mu$m bands 
\citep[e.g.,][]{fazio2004}.
PAHs are typically fluorescing due to the
absorption of ultraviolet photons from the surrounding interstellar radiation field;
thus, PAH emission is a ubiquitous feature of the diffuse interstellar 
medium \citep[e.g.,][]{flagey2006}.
So while the proplyds in Orion are seen against the bright nebula
emission, L1157 is seen against PAH emission.  The L1157 absorption feature
is large, $\sim$1-2$\arcmin$ or $\sim$15,000-30,000 AU at a distance of 250 pc
(see \S \ref{dist}) and flattened with an axis ratio of $\sim$0.25.
Due to the size and lack of any kinematic information, 
we will call it a flattened envelope or a pseudo-disk, not a circumstellar disk;
such flattened density enhancements are expected from many theoretical
constructs that include magnetic fields or rotation in the collapse process
\citep[e.g.,][]{tsc1984,fiedler1993, galli1993a, galli1993b,
hartmann1994, hartmann1996}.

The observed extinction is probably
dominated by a combination of silicates and ice.
The 8.0 $\mu$m channel overlaps
with the 9.7 $\mu$m silicate feature in half of the bandpass. However,
ice features are present in both the 5.8 and 8.0 $\mu$m channels: the 6.85
$\mu$m (CH$_3$OH or NH$^-_4$) and the 9.0 $\mu$m NH$_3$ ice features 
in the 8.0 $\mu$m channel, and the 6.0 $\mu$m H$_2$O ice feature
in the 5.8 $\mu$m channel. 
The qualitative appearance of the absorption does correlate very well with observations
of the dust continuum emission structure seen at $\lambda$~=~1.3~mm \citep{gueth2003},
as well as the ammonia emission \citet{bachiller1993}, but as these
trace the dense core, this may not be especially surprising.

\subsection{Outflow Features}

L1157 has one of the most studied and well-developed bipolar outflows.
The outflow has an inclination of $\sim$80$\arcdeg$ and a slower
blueshifted (southern) lobe than the redshifted (northern) lobe
\citep[e.g.][]{gueth1996,bachiller2001}.  
The S shape morphology
and the three peaks at point reflection symmetry seen in CO
and SiO emission are well explained by outflow 
precession with a cone angle of 
$\sim$15$\arcdeg$ \citep{gueth1996,zhang2000,bachiller2001}.

The IRAC data have a remarkable coincidence of emission structure with the molecular outflows,
Figure \ref{disk}.
We
measure the largest separation of peaks as $\sim$15$\arcdeg$ and an angle
of the outflow extension in width (``east-west'' direction) as
$\sim$35$\arcdeg$.  This means that the precession cone has a
$\sim$15$\arcdeg$ angle, consistent with previous studies
\citep[e.g.][]{zhang2000,bachiller2001}, and that each episodic shock
has $\sim$10$\arcdeg$ opening angle.  The precession cone angle here
indicates the total angle of the cone, and the opening angle represents
half of the outflow opening.  From the peak positions, the precession
period is estimated as $\sim$3050 years, assuming 250 pc distance and
100 km s$^{-1}$ constant outflow velocity from the model of \citet{bachiller2001}.

Unlike the extinction structure, 
the outflow features are commonly shown in all four
IRAC bands.
The excitation
mechanism for the broad energy range of emission is beyond the scope of this letter; 
it requires spectroscopic observations
and detailed modeling to understand level populations in a large energy
region.  However, 
we can consider hydrogen recombination lines 
\citep[the bipolar outflow regions were dissociatively shocked, e.g.,][]{bachiller2001}, 
and molecular hydrogen ro-vibrational and rotational lines (based on the chemical complexity
of the outflow).
In fact, many of these transitions 
are in the IRAC bands with some metal ion fine
structure lines \citep[e.g.][]{dirk2000,noriega2004,neufeld2006}.  The
3.6 $\mu$m band may be dominated by H$_2$ 1-0 O(5,6,7) and 0-0 S(13). 
IRAC bands 2, 3, and 4 may be dominated by H$_2$ 0-0 S(12,11,10,9),
S(8,7,6), and S(5,4), respectively. Some hydrogen recombination lines
such as Brackett $\alpha$ in band 2 and ion fine structures such as [Ni II],
[Ar II], and [Ar III] in band 4 also may contribute.   Finally, PAH
emission contributes in bands 3 and 4.

\subsection{Optical Depth Measurement and Models}

The absorption feature of the flattened envelope structure in L1157
provides an excellent opportunity to examine the envelope material
by measuring the optical depth along the feature.  This is especially true
as the L1157 flattened envelope
is nearly edge-on, isolated, and 
surrounded by relatively smooth background emission.  The background
and associated average uncertainty was measured using the IRAF\footnotemark\ task
``imstat'' in three areas within 90$\arcsec$ from the central
source (0.54 $\pm$ 0.04 MJy/sr).  The areas were free of stars and not significantly affected by
residual absorption from the outer envelope. A constant background
may not be entirely realistic, but we did explore the possibility of modeling the background
using median filtering in a similar method to \citet{simon2006}. However,
using a background model did not improve the analysis.

\footnotetext{IRAF is distributed by the National Optical Astronomy Observatories,
    which are operated by the Association of Universities for Research
    in Astronomy, Inc., under cooperative agreement with the National
    Science Foundation.}

The intensity of the
absorption feature was measured perpendicular to the outflow, radially
away from the central source.  This was done using SAOImage DS9 to measure
a ``projection'' of 82 pixels in length and 3 pixels in width.  
The intensity was taken as the average of the
3 pixels (in width) for a particular position along the projection.
To determine the intensity uncertainty, we used the pixel values
provided by the uncertainty frame from the \textit{Spitzer} pipeline.
Note, that the point spread function of IRAC at 8.0$\mu$m is 
1$\farcs$9, and the pixel size is 1$\farcs$2 \citep{fazio2004}.

In order to quantitatively compare the absorption to axisymmetric
models of circumstellar envelopes, we created a simple model of the
opacity through an absorption slice, perpendicular to the outflow.
The model consists of an edge-on flattened object, i.e. disk-like,
with a radial density profile, $\rho(r) = \rho_0 (r/r_0)^{-p}$,
where $r_0$ and $\rho_0$ are the radius and the density one pixel away from
the center.
Using a Cartesian convention where location along the absorption
feature is $x$ and the distance along the line of sight is $y$, one can
re-write the density profile along the absorption as $\rho(x,y)= \rho_0
((x^2+y^2)/r^2_0)^{-p/2}$ The opacity is then of the form
\begin{equation}
\tau(x)~=~2\kappa_\lambda \rho_0 r_0 \int ^{\sqrt{R^2-x^2}}_0 ((x^2+y^2)/r^2_0)^{-p/2}d(y/r_0),
\end{equation}
where R is the outer radius of the disk and $\kappa_\lambda$ is the dust opacity.   
This is compared to the
observed opacity of $\tau(x) = - ln({\rm intensity(x)/background})$ in the average
along the projection.  This is an edge-on approximation; the source
has an $\sim$80$\arcdeg$ inclination, but as we are averaging over three vertical
pixels, the effect is minimized.

To fit this model to the data, a grid of models was used for the
following variables: the background level (although we measure a background level, we still
use a grid of values around the measured value), a constant for $\kappa_\lambda$ and $\rho_0$
combined, 
the outer disk radius ($R$), the location of the structure center in pixels ($X_0$),
and the power law dependence
($p$). Note that there are no {\it a priori} assumptions as to the
values of $\kappa_\lambda$ or $\rho_0$. 
This parameter space is then
compared to the 8.0 $\mu$m data using a $\chi^2$ likelihood.
The pixels close to the central source were not used in the fits, as they contain
bright emission from the scattered light close to the protostar, see Figure \ref{fit}.

Our model parameter grid used background levels = 0.45 to 0.60 MJy/sr in steps of
0.01, p=0.5 to 3 in steps of 0.5, $R$=18$\arcsec$ to 84$\arcsec$ in steps of 1.2$\arcsec$,
$X_0$ was fixed to 0$\arcsec$ or $\pm$1.2$\arcsec$ of the peak of the compact
dust emission \citep{beltran2004},
and the constant=0.01 to 2.0 in steps of 0.01.
The model had nearly 3 million grid points.
Using these parameters, the simple model
of optical depth successfully fit the data with high
confidence levels. Models were considered good fits if the confidence level was $>$90$\%$, i.e.
$>$10$\%$ likelihood (see Table \ref{fits}).
In general, the fits are not well constrained.
The power-law index, $p$, fits range from 0.5 to 2.0, and the outer radius, $R$, was constrained
to $\ge$27.6$\arcsec$; however, the maximum is not well 
constrained as we only modeled $\pm$50$\arcsec$
of data.
Examples of the ``best fits'' for each acceptable density power-law are shown 
on the 8.0 $\mu$m data in Figure \ref{fit}.  

\section{Discussion}

The shape of the absorption feature is especially intriguing, as it looks
like a disk structure perpendicular to the outflow axis.
This is the first clear detection of a flattened envelope or pseudo-disk in
a Class 0 object.
\cite{galli1993a,galli1993b} have shown that a modest magnetic field structure
modifies infall from the initial spherical cloud to form a so-called
``pseudo-disk''; a flat thin structure in the equatorial
plane that is not rotationally-supported, thus collapsing.
This type of structure is also seen from simple flat sheet models of
collapse \citep[e.g.,][]{hartmann1994,hartmann1996}, as well as detailed
ambipolar diffusion models \citep[e.g.,][]{fiedler1993}.
On the other hand, this structure is large $\sim$15,000-30,000 AU, depending
on the background used.
That size is somewhat larger than
the inner envelope size estimated from interferometric models of the
dust continuum \citep{looney2003}.  However, the single-dish dust emission 
\citep{gueth2003} is extended along the same axis as the absorption, which
argues that the inner envelope in the equatorial plane
either has higher density, and/or different dust opacity properties.

Our modeling results show that the properties of the structure, as determined
by the absorption model, are consistent with the above theoretical constructs,
i.e. flattened envelopes and density profiles. 
Although we model a range of indexes ranging from p = 0.5 to 3, only
the 0.5 to 2 provide acceptable fits at the 90\% confidence level
with the vast majority of fits being p=1.5.  

To better explore the physical meaning the model, we assume
dust opacities (dust plus gas)
from \citet{lidraine2001} of $\kappa_{8.0\mu m} =5.912~cm^2~g^{-1}$.
Although using interstellar dust opacities for $\kappa_\lambda$ is
probably incorrect, as Class 0 sources are thought to have already
experienced some grain growth \citep[e.g.,][]{looney2003,natta2007},
it is still a useful approximation.  Using the assumed $\kappa_{8.0\mu
m}$, the derived range for the density reference, $\rho_0$, or 1 pixel
(1$\farcs$2) from the center of the envelope (i.e. 300 AU at 250 pc)
is listed in Table \ref{fits}.

\begin{figure}
\includegraphics{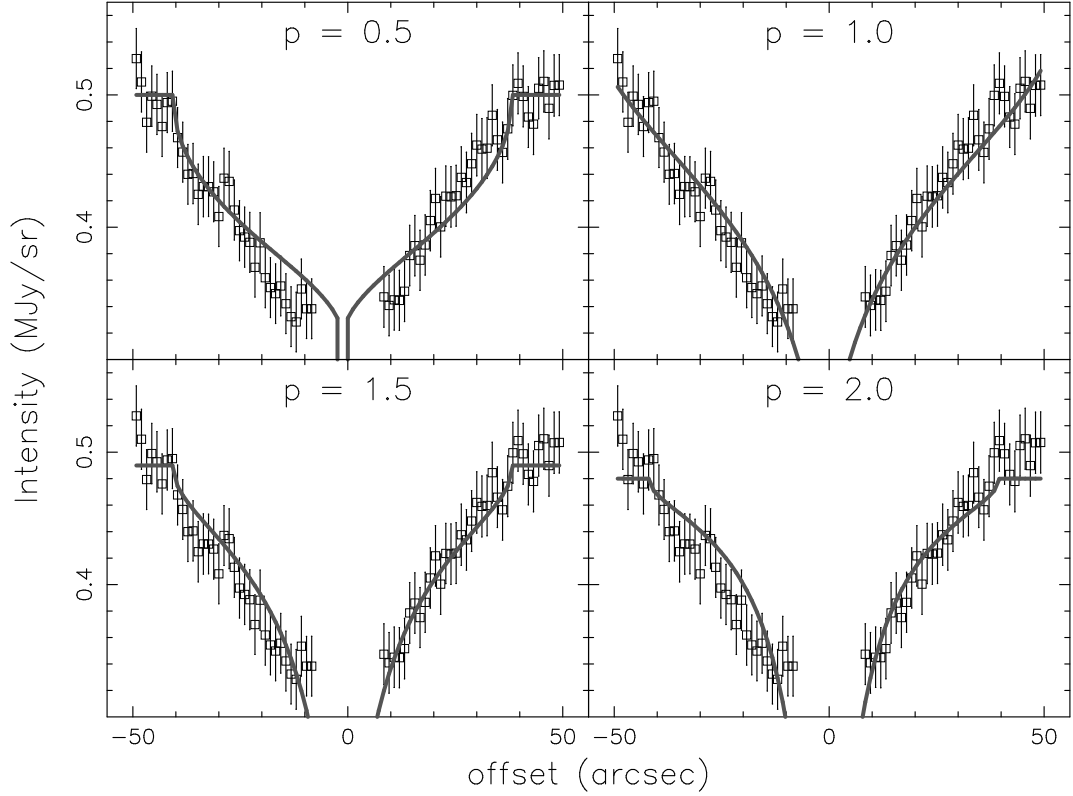}
\vspace{8cm}
\caption{``Best fit'' examples for the 4 acceptable density power-law indexes on the
8 $\mu$m data.  The fit parameters are: 
(p=0.5) background = 0.5 MJy/sr, constant =  0.02, $X_0$ = -1.2$\arcsec$,
and R = 39.6$\arcsec$;
(p=1) background = 0.57 MJy/sr, constant = 0.11,
X$_0$ = -1.2$\arcsec$, and R = 55.2$\arcsec$; (p=1.5) background = 0.49 MJy/sr, 
constant = 0.37, X$_0$ = -1.2$\arcsec$,
and R = 55.2$\arcsec$; (p=2.0) background = 0.48 MJy/sr, constant = 1.3, X$_0$ = -1.2$\arcsec$,
and R = 40.8$\arcsec$.
}
\label{fit}
\end{figure}

In addition, we can estimate the absorbing mass of the flattened envelope component
for each ``best fit'' model of Figure \ref{fit} and 
a height of 3 pixels (the size of the box we averaged over).
We assume that the vertical density profile is constant for the mass estimate, even though
the observed absorption falls off vertically with scale heights of $\sim$3-4 pixels, using a Gaussian
vertical structure.
Our mass estimate, listed in Table \ref{fits}, ranges from 0.08 to 0.16 M$_\odot$.
Without using any model, we can also estimate the mass required for the observed 
extinction.  
We use the above value for $\kappa_{8.0\mu m}$ and a background of 0.54 MJy/sr
to calculate the mass necessary for the extinction of all pixels in the central
region below 0.458 MJy/sr (577 pixels).
The total absorbing mass required for those pixels is 0.19 M$_{\odot}$.
It is important to note that this mass is only in the absorbing pseudo-disk, 
but the mass is comparable to 
the 140$\arcsec$ extended envelope detected in the millimeter continuum
as discussed in \citet{gueth2003} with an estimated mass
of 0.7 M$_{\odot}$ at an assumed distance of 250 pc.  This implies that although a
large fraction of the mass of the envelope is in the flattened structure, most
of the mass is more diffuse.
Due to the log nature of the mass absorption, the density contrast
from the center of the absorption feature to slightly offset from the absorption
feature is approximately an order of magnitude.
This is more than expected from the numerical models of \cite{galli1993b}, but somewhat
consistent with the ambipolar models of \cite{fiedler1993} and the sheet collapse model
of \cite{hartmann1996}.

As can be seen in Table \ref{fits} and Figure \ref{fit}, these data can not 
constrain the model density profiles directly; 
multiple power-laws are allowed.  On the other hand,
although many theoretical models do suggest a flattened envelope structure, the
power-law index in the pseudo-disk is expected to evolve.  In that case, we would not
expect a single power-law to well describe the envelope density.
In fact, the ambipolar models \citep[e.g.][]{tassis2005a,tassis2005b} suggest that
the power-law can be episodic in the flattened envelope.
Further studies with increased sensitivity need to be compared directly to the
theoretical density profiles in the flattened objects to say anything more completely.

\acknowledgements
We thank F. Gueth for providing the $\lambda$~=~1.3~mm continuum emission map
and R. Bachiller for providing the CO 2-1 emission map.
This work is based on observations made with the Spitzer Space Telescope,
which is operated by the Jet Propulsion Laboratory, California Institute
of Technology under a contract with NASA. Support for this work was
provided by NASA.
This research has made use of SAOImage DS9, developed by Smithsonian Astrophysical Observatory

\bibliographystyle{apj}
\bibliography{ms}
\clearpage

\begin{deluxetable}{cccccccc}
\tabletypesize{\scriptsize} 
\tablewidth{0pt}
\tablecaption{Summary of Acceptable Fits}
\tablehead{
  \colhead{Power law} & \colhead{Background} & \colhead{$R$} &
  \colhead{Constant} &  
  \colhead{$\rho_0$\tablenotemark{a}} & \colhead{Mass\tablenotemark{b}} &
  \colhead{Max.} & \colhead{Number} \\
  \colhead{p} & \colhead{(MJy/sr)} & \colhead{($\arcsec$)} & 
  \colhead{($\kappa_\lambda \rho_0 r_0$)} & 
  \colhead{($10^{-18}$ $g/cm^3$)} & \colhead{(M$_\odot$)} &
  \colhead{prob. (\%)} & \colhead{of fits} \\
  }
\startdata
0.5 & 0.47 - 0.60 & 27.6 - 50.4 & 0.02 - 0.03 & 0.75 - 1.13   & 0.08 & 99.2 & 130 \\
1.0 & 0.46 - 0.60 & 28.8 - 84.0 & 0.06 - 0.13 & 2.26 - 4.90   & 0.16 & 99.9 & 1820 \\
1.5 & 0.46 - 0.60 & 30.0 - 84.0 & 0.26 - 0.56 & 9.80 - 21.11  & 0.14 & 99.9 & 7253 \\
2.0 & 0.47 - 0.51 & 34.8 - 84.0 & 1.04 - 1.63 & 39.20 - 61.43 & 0.15 & 48.3 & 2622 \\
\enddata
\label{fits}
\tablenotetext{a}{Assuming d=250 pc, for other distances multiply by (250/d).}
\tablenotetext{b}{The mass is estimated using the ``best-fit'' models from Figure \ref{fit}
with d=250 pc.  For other distances multiply by (d/250)$^3$.}
\end{deluxetable}
\clearpage

\clearpage

\clearpage

\end{document}